\documentclass[comsoc,10pt]{IEEEtran}
\usepackage[T1]{fontenc} 
\usepackage[utf8]{inputenc}
\usepackage{amsmath}
\usepackage[cmintegrals]{newtxmath}
\usepackage{bm} 
\usepackage{siunitx}
\usepackage{tikz}
\usetikzlibrary{angles, quotes}
\usepackage{tikz-imagelabels}
\usepackage{pgfplots}
\usepackage{subcaption}

\usepackage{fancyhdr}
\usepackage{todonotes}
\pgfplotsset{compat=1.6}
\usetikzlibrary{shapes, arrows, arrows.meta, graphs, decorations,decorations.markings,decorations.shapes}
\usepackage{pgfplots}
\pgfplotsset{compat=1.12}
\usepgfplotslibrary{external,polar}
\usepackage{soul}
\usepackage[draft,xcolor={red,dvpipdf}]{changes}
\newlength{\figW}
\newlength{\figH}
\definechangesauthor[name={Nils Johannsen}, color=blue]{NJ}
\definechangesauthor[name={Lukas Grundmann}, color=red]{LG}
\definechangesauthor[name={Max Schurwanz}, color=green]{MASC}

\colorlet{Changes@Color}{red}

\usepackage{mathrsfs}
\graphicspath{{./pic/}}
\def\BibTeX{{\rm B\kern-.05em{\sc i\kern-.025em b}\kern-.08em
		T\kern-.1667em\lower.7ex\hbox{E}\kern-.125emX}}

\IEEEoverridecommandlockouts

\makeatletter
\newcommand*\titleheader[1]{\gdef\@titleheader{#1}}
\AtBeginDocument{%
	\let\st@red@title\@title
	\def\@title{%
		\bgroup\normalfont\large\centering\@titleheader\par\egroup
		\vskip1.5em\st@red@title}
}
\makeatother

	\title{Joint Communication, Sensing and Localization for Airborne Applications}
			\titleheader{This work has been submitted to the IEEE for possible publication. Copyright may be transferred without notice, after which this version may no longer be accessible.}

\author{
	\IEEEauthorblockN{Nils L. Johannsen$^{(1)}$, Max Schurwanz$^{(2)}$, Lukas Grundmann$^{(3)}$, \IEEEmembership{Graduate Student Member, IEEE}, Jan~Mietzner$^{(2)}$, \IEEEmembership{Senior Member, IEEE}, Dirk Manteuffel$^{(3)}$, \IEEEmembership{Member, IEEE},  and Peter A. Hoeher$^{(1)}$, \IEEEmembership{Fellow, IEEE}}%
 \thanks{The authors are with the:
	\\$^{(1)}$ \textit{Chair of Information and Coding Theory}, \textit{Kiel University}, Kiel, Germany, email: \{nj, ph\}@tf.uni-kiel.de
	\\$^{(2)}$ \textit{Department of Media Technology}, \textit{University of Applied Sciences (HAW) Hamburg}, Hamburg, Germany, email: \{Max.Schurwanz, Jan.Mietzner\}@haw-hamburg.de
	\\$^{(3)}$ \textit{Institute of Wireless and Microwave Systems}, \textit{Leibniz University Hannover}, Hannover, Germany, email: \{grundmann, manteuffel\}@imw.uni-hannover.de}
}

\usepackage{environ}
\makeatletter
\newsavebox{\measure@tikzpicture}
\NewEnviron{scaletikzpicturetowidth}[1]{%
	\def\tikz@width{#1}%
	\begin{lrbox}{\measure@tikzpicture}%
		\BODY
	\end{lrbox}%
	\pgfmathparse{#1/\wd\measure@tikzpicture}%
	\BODY
}
\makeatother

\begin{document}
	\pgfplotsset{compat=1.3,
		every axis legend/.style={
			y tick label style={/pgf/number format/1000 sep=},					
			x tick label style={/pgf/number format/1000 sep=},
			z tick label style={/pgf/number format/1000 sep=}
	}}

	\maketitle
	\begin{abstract}
	    With the upcoming trends in autonomous driving and urban air mobility, the number of self-navigating vehicles will increase, since they are foreseen for deliveries as well as autonomous taxis among other applications. 
	    To this end, a multitude of on-board systems for wireless communication, environment sensing, and localization will become mandatory.
	    This is particularly true for unmanned aerial vehicles (UAVs), since participation in the airspace requires compatibility to and safe interaction with established users. A certain number of systems are already in-use and occupy defined spectra as well as installation space, which limits the freedom in the design of new systems. 
	    The miniaturization of aerial vehicles like drones for delivery services further reduces the degrees of freedom, especially in terms of size and weight of any additional equipment.
	    Hence, in this paper a joint approach of the design of joint communication, sensing and localization for UAVs is discussed.
	    Towards this goal, multi-mode multi-port antennas and joint waveform design are proposed as a part of the solution, when elevating autonomous driving to the third dimension.  
	\end{abstract}
	

	\section{Introduction}
	\label{sec:Introduction}
	Enabled by recent technologies and large demands in fast delivery and transport, the density of aerial vehicles will be increased drastically in the airspace.
	Especially when operating unmanned aerial vehicles (UAVs), safety is mandatory as a first prerequisite.
	The improvement of safety has always been a major target in research and development of new radio-based systems for aerial applications.
	These systems include use for communication, navigation, and collision avoidance~\cite{NLALO13,PTSTJ18,BDOACS21,ED275}.
    Several radio-based applications are in use in general aviation, some of which cooperatively exchange information with other airspace participants, while others non-cooperatively collect information about the environment only.
	In recent developments, the visible spectrum is taken into account as well, e.g. by including cameras.
	However, focus of this article is on the option of combining different radio applications in a single system.
	A simple example of multiple use of signals in airborne applications is the advanced collision avoidance system (ACAS).
	Initially, transponders have been developed for military applications to allow the identification of friend or foe (IFF), introducing a secondary radar.
	Today, transponders allow the flight controllers an improved situational awareness by providing flight altitude and further information~\cite{ED73C}, while enabling collision avoidance based on the mutual exchange and interrogation of transponder data~\cite{ED275}.
    
    The combination of communication and sensor technology is of great interest, as it opens up further degrees of freedom in the processing of the information obtained.
    Joint communication, sensing and localization (JCSL) can provide improved situational awareness and reliability in the development of small, agile, and customized transportation solutions. 
    Data rates can be increased and the number of independent radio-frequency (RF) systems that need to be individually coordinated for simultaneous use can be reduced.
    The ongoing Master360 project (``Multisensor System for Helicopter Automation and Safe Integration of UAVs into Air Traffic with 360$^\circ$ Coverage'') is dedicated to safety issues of UAVs with particular interest on unmanned autonomous flying taxis.
    The goal of the project is the safe integration of UAVs in the segregated airspace.
    A wide variety of on-board communication and sensing systems must be coordinated to enable reliable operation of the aerial platforms.
    Therefore, a new chipset for the radar signal processing is developed which allows the miniaturization of the radar system.
    The novelties mentioned in this paper are exemplified by the developments in the Master360 project, illustrating possible benefits through JCSL developments.
    
    Novel aspects of the paper include the following:
    \begin{itemize}
        \item Scenarios for implementing JCSL in urban air mobility (UAM) settings and aerial applications are presented and illustrated with practical examples,
        \item a waveform design for JCSL applications is proposed, and
        \item multi-mode multi-port antennas are shown to provide good (beamforming) performance to achieve the desired goals.
    \end{itemize}
    The remainder is organized as follows: In Section~\ref{sec:ClassificationCAU}, different air-based communication applications and their origins are discussed and the potential regarding JCSL is outlined.
    In Section~\ref{sec:SensingHAW}, the increasing demand in sensing and localization applications in aviation is pointed out.
    The challenges arising from the application example in the project motivate a joint waveform design for JCSL introduced in Section~\ref{sec:WaveformDesignHAW}.
    In Section~\ref{sec:AntennasLUH}, multi-mode multi-port antennas (M$^3$PAs) are proposed as a key-feature to solve demanding tasks.
    Finally, conclusions are drawn in Section~\ref{sec:Conclusion}.

	\section{Classification of Radio Communication Applications in Aviation}
	\label{sec:ClassificationCAU}
	Beginning with the early days of aviation, communication applications in aviation have been of special interest to both flight crews and ground personnel. 
    In addition to the early stages of transponders for IFF mentioned in the introduction, British aircraft were equipped with wireless telegraphy as early as World War I to allow rapid transmission of information from tracking pilots to command~\cite{Mar20}.
	Both communication systems worked in different manner, serving different tasks: communication in terms of a human to human (H2H) connection (radio application) and human to system / machine (H2M).
	Due to the different nature of the communication (H2H vs. H2M), different techniques were used.
	
	This has not changed even for the latest versions of radio systems aboard airplanes.
	In general, most systems have been developed to solve a single task.
	Some of them have been used for additional tasks compared to what they have initially been developed for.
    ACAS II, for example, which provides advice on how to avoid hazardous situations in air traffic, was derived from the use of transponder signals, which were originally introduced to provide controllers with better situational awareness.
	When trying to improve JCSL, the different kinds of communication services aboard an aerial vehicle need to be distinguished and compared.
	Several options of comparison are available: Some systems are based on the ground, whereas others are employed airborne.
	Some transmit voice, using analog modulation techniques, while others transmit data using digital schemes.
	In the context of this paper, the communication services are sorted by their use-case:
	\begin{itemize}
	    \item Telephony,
	    \item collision avoidance, and
	    \item navigation.
	\end{itemize}
	
	The use-case of the communication technology determines required bandwidth and signal design, including modulation and transmit power.
    The signal design is particularly important for the implementation of JCSL, as the applications must not have a negative impact on the coexisting systems.
	\begin{figure}
	    \centering
	    \input{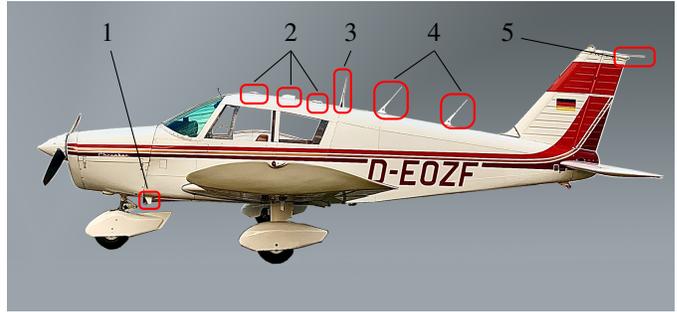}
	    \caption{Piper with antennas for different purposes. The numbers indicate functions and typical frequencies as follows: 1: Transponder (1030/1090 MHz), 2: GNSS / SatCom (GPS for aerial applications: 1176.45 MHz), 3: ELT (406 MHz), 4: Radio communication (117.975 - 137 MHz), 5: VOR navigation (108 - 117.95 MHz).}
	    \label{fig: PiperAntennas}
	\end{figure}
	In Fig.~\ref{fig: PiperAntennas}, a Piper plane with its antennas is shown.
    The antennas can be related to the applications aboard the plane.
	The probably best-known application is radio telephony. 
	Using analog amplitude modulation in the frequency band between 117.975~\si{MHz} and 137~\si{MHz} allows for understanding of speech even under poor signal-to-noise ratio (SNR) conditions, which would lead to a connection loss using alternative transmission schemes.
	The channels for radio telephony are separated by 8.33~\si{kHz}.
	
	Flight transponders usually have to be seen as a part of the secondary radar, providing flight controllers required information to allow safe travel.
	Initially, in the simple modes supported by early transponders, only a given code, named "squawk", was transmitted. 
	However, nowadays these transponders are capable of providing a 24 bit address of the aircraft, altitude and velocity, and additional information which either include rate of climb or descent, and allow the use for collision avoidance.
    This is based on interrogating transponders from other aircraft.
	Corresponding transponders being capable of interacting with other transponders in such a way are called interrogators.
	
	Apart from the ACAS~II system, various additional collision avoidance systems are in widespread use, depending on the application.
    In terms of unmanned aircraft integration, ACAS~X is a new system rather than an improvement on the established ACAS~II.
	It includes communication with ACAS~II employed aerial vehicles, but offers new options.
	The new ACAS~X standard family includes ACAS~Xu for unmanned and ACAS~Xa for large aircraft~\cite{MaJe16}. 
    For small aircraft, including gliders, the FLARM system (868~- 915~MHz) has been developed to assist the pilot to prevent collisions e.g. in small lifts and at rather short distances.
    This system does not provide any resolution advisory, but provides information about the conflict with highest probability.
    If a collision of the aircraft happens, the emergency locator transmitter (ELT) on motorized airplanes starts its transmission due to either acceleration, contact to water, or manual activation.
    The device transmits a locator signal at 121.5~MHz and 406~MHz.
    The locator signal contains identification data and allows a localization based on e.g. Doppler shift-based algorithms.
    
	Employing the same frequency band as radio telephony, but using a spacing of 50~\si{kHz}, very high frequency (VHF) omni-directional radio range (VOR) systems allow navigation without Global Navigation Satellite Systems (GNSS) like the Global Positioning System (GPS).
	The signal of the ground stations contains the angle of the aerial vehicle measured w.r.t. magnetic north, modulated as a Doppler frequency shift.
	Improved versions also provide the option of distance measuring, using distance measurement equipment (DME).
	These systems work similar to a secondary radar: The vehicle transmits a number of pulses, which are then answered by the ground station.
	Since the delay of the signal processing in the ground station is defined, the processor in the receiver aboard can calculate the time of flight of the signals, the so-called round-trip time.
	This may serve as an example of a narrow line between localization and communication in aviation applications.
	The DME system works by utilizing a kind of code division multiple access (CDMA) scheme: The pulses are transmitted in a randomly chosen manner to distinguish the pattern of replies between the own call and others.
	An extensive literature survey on navigation for aerial vehicles is given in \cite{CaPeAl18}.
	Even older than VOR-DME is the usage of non-directional beacons (NDBs), which can be seen as an equivalent to the naval lighthouses for guiding planes.
    These old navigation aids provide a low-frequency identification signal only, of which the direction-of-arrival (DoA) needs to be measured or estimated, using a combination of directional and non-directional antennas, denoted automatic direction finder (ADF).
    Nowadays, most navigation is done or assisted by GPS.
    However, as reported in \cite{Har21}, relying on a single navigation aid like GNSS is not sufficient due to possible system failure (single point of failure). 

	In summary, the above mentioned communication systems have traditionally been treated separately. 
	They employ their own antennas as well as individual installation space and signal processing units and occupy different frequency bands. 
	

	\section{Radar Sensing and Localization in Aviation}
    \label{sec:SensingHAW}
    When Christian Hülsmeyer filed the first radar patent in 1904 to detect ships in all weather conditions - especially fog - the first radar application was invented.
    During the Second World War, the focus shifted towards ground-based applications for detection (sensing) and localization of aerial vehicles, for example for early-warning and fire-control tasks.
    After the World War, the compiled radar knowledge migrated into various civil applications including ground-based radars for air-traffic control and airborne radar systems for pilot assistance and situational awareness.
    Nowadays, weather radar and altimetry radar are mandatory systems to be installed in large airplanes.
    Compared to optical sensors (e.g. cameras), a radar system offers all-weather capabilities and day-and-night operation.
    Recent trends are tending towards UAM, and new flight maneuvers in aviation, especially in the vertical dimension during automatic / vertical take-off and landing (ATOL / VTOL) as well as en-route demand for fast, secure, light-weight, cost-effective and small installation-size radar systems with basically 360$^\circ$ coverage in azimuth and elevation.

	Due to the densification of traffic patterns especially expected for UAM scenarios, aerial vehicles will be required to be equipped with multiple systems capable of handling radio-based communication and localization / sensing tasks both automatically and reliably. 
	If realized as separate systems, confined installation spaces are likely to entail increased levels of interference leading to basic coexistence issues.
	Mutual system design may alleviate these problems by either utilizing the frequency or time domain via advanced signal processing (Section~\ref{sec:WaveformDesignHAW}), the spatial domain via intelligent hardware design (Section~\ref{sec:AntennasLUH}), or a combination thereof.

    \section{Waveform Design for JCSL in Aviation}
    \label{sec:WaveformDesignHAW}
    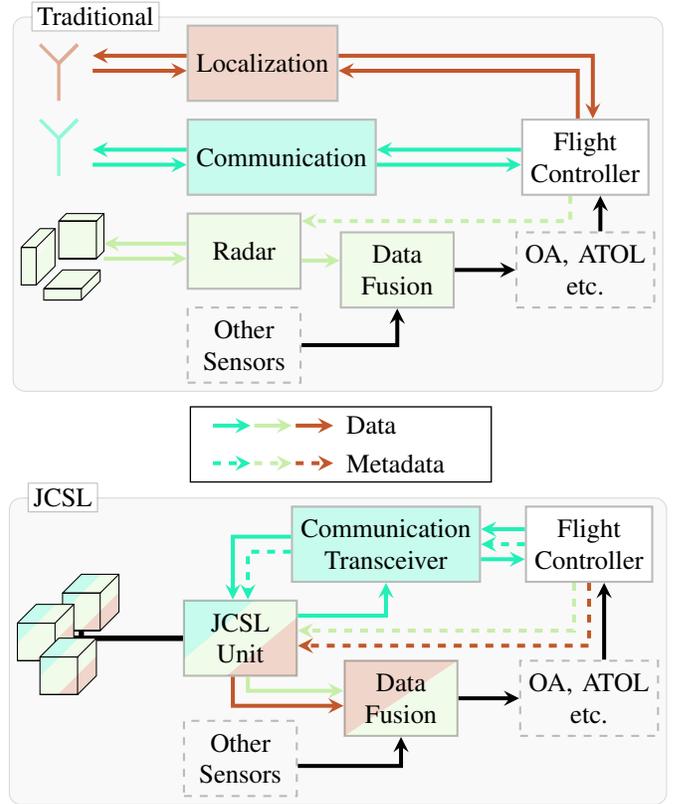
\begin{figure}
        \centering
        \usetikzlibrary{calc}                   
\usetikzlibrary{shapes}      
\usetikzlibrary{arrows,arrows.meta,decorations.pathmorphing,backgrounds,positioning,fit,petri,patterns}
\usetikzlibrary{shapes.geometric,backgrounds}

\tikzset{
  mybackground/.style={execute at end picture={
        \begin{scope}[on background layer]
          \draw[black!15,fill=gray!5,rounded corners=1ex] ($(current bounding box.south west)+(-0.1,-0.1)$) rectangle ($(current bounding box.north east)+(0.1,0.1)$);
          \node[draw,fill=white,rectangle,rounded corners=0.1,anchor=west,inner sep=2pt,minimum width=4ex,draw opacity=0.25] at ($(current bounding box.north west)+(0.25,0)$) {#1};
        \end{scope}
    }},
}

\begin{tikzpicture}[mybackground={Traditional}]
\definecolor{commcolor}{RGB}{33,240,182};
\definecolor{loccolor}{RGB}{194,87,43};
\definecolor{radcolor}{RGB}{199,238,168};

\tikzstyle{block} = [thick,align=center,minimum width=1.5cm,minimum height=1cm,draw,color=black,draw opacity=0.25]
\tikzstyle{antenna} = [black,draw,minimum width=0.5cm,minimum height=0.5cm]
\tikzstyle{darrow} = [line width=1.5pt,-stealth]
\tikzstyle{commarrow} = [darrow,color=commcolor]
\tikzstyle{locarrow} = [darrow,color=loccolor]
\tikzstyle{radarrow} = [darrow,color=radcolor]
\tikzset{
diagonal fill/.style 2 args={fill=#2, path picture={
    \fill[#1, sharp corners] (path picture bounding box.south west) -|
                         (path picture bounding box.north east) -- cycle;}},
three stripes fill/.style n args={3}{fill=#2, path picture={
    \fill[#1, sharp corners] (path picture bounding box.70) -| (path picture bounding box.200) -- cycle; 
    \fill[#3, sharp corners] (path picture bounding box.250) -| (path picture bounding box.20) -- cycle;}}
}
    \coordinate (loc1) at(-2,2);
    \coordinate (comm1) at (-2,1);
    \coordinate (rad1) at (-2,-0.1);
    \coordinate (rad2) at (-2.5,-0.45);
    \coordinate (rad3) at (-2.2,-0.65);

    \node[block,fill=radcolor!25,anchor=west] (signalfusion) at (-0.3,0) {Radar};
    \node[block,fill=commcolor!25,anchor=west] (commtrans) at (-0.3,1.25) {Communication};
    \node[block,fill=loccolor!25,anchor=west] (loctrans) at (-0.3,2.5) {Localization};
    \node[block,fill=radcolor!25] (datafusion) at (2.5,-0.25) {Data\\Fusion};
    
    \node[block,dashed,anchor=west] (camera) at (-0.3,-1.25) {Other\\Sensors};
    \node[block,dashed] (proc) at (5,-0.25) {OA, ATOL\\etc.};

    \node[block,fill=white] (mainproc) at (5,1.25) {Flight\\Controller};

    \draw[antenna,line width=1.25pt,color=loccolor!50] (loc1) -- ++(0,0.5) -- ++(0.25,0.25) -- ++(-0.25,-0.25) -- ++(-0.25,0.25);

    \draw[antenna,line width=1.25pt,color=commcolor!50] (comm1) -- ++(0,0.5) -- ++(0.25,0.25) -- ++(-0.25,-0.25) -- ++(-0.25,0.25);
    
    \draw[antenna,fill=radcolor!25] (rad1) -- ++(0.5,0) -- ++(0.1,0.1) -- ++(0,0.5) -- ++(-0.5,0) -- ++(-0.1,-0.1) -- ++(0,-0.5) -- ++(0.5,0) -- ++(0,0.5) -- ++(-0.5,0) -- ++(0.5,0) -- ++(0.1,0.1);
    \draw[antenna,fill=radcolor!25] (rad2) -- ++(0.2,0) -- ++(0.25,0.25) -- ++(0,0.5) -- ++(-0.2,0) -- ++(-0.25,-0.25) -- ++(0,-0.5) -- ++(0.2,0) -- ++(0,0.5) -- ++(-0.2,0) -- ++(0.2,0) -- ++(0.25,0.25);
    \draw[antenna,fill=radcolor!25] (rad3) -- ++(0.5,0) -- ++(0.25,0.25) -- ++(0,0.15) -- ++(-0.5,0) -- ++(-0.25,-0.25) -- ++(0,-0.15) -- ++(0.5,0) -- ++(0,0.15) -- ++(-0.5,0) -- ++(0.5,0) -- ++(0.25,0.25);

    \draw[locarrow] ($(loctrans.west)+(0,0.1)$) -- ($(loc1)+(0.45,0.6)$);
    \draw[locarrow] ($(loc1)+(0.45,0.4)$) -- ($(loctrans.west)+(0,-0.1)$);
    
    \draw[commarrow] ($(commtrans.west)+(0,0.1)$) -- ($(comm1)+(0.45,0.35)$);
    \draw[commarrow] ($(comm1)+(0.45,0.15)$) -- ($(commtrans.west)+(0,-0.1)$);

    \draw[radarrow] ($(signalfusion.west)+(0,0.1)$) -- ($(rad1)+(0.6,0.2)$);
    \draw[radarrow] ($(rad1)+(0.6,0)$) -- ($(signalfusion.west)+(0,-0.1)$);

    \draw[commarrow] ($(mainproc.west)+(0,0.1)$) -- ($(commtrans.east)+(0,0.1)$);
    \draw[commarrow] ($(commtrans.east)+(0,-0.1)$) -- ($(mainproc.west)+(0,-0.1)$);
    \draw[radarrow,dashed] ($(mainproc.south)+(-0.2,0)$) |- ($(signalfusion.east)+(0,0.4)$);
    \draw[radarrow] ($(signalfusion.east)+(0,-0.125)$) -- ($(datafusion.west)+(0,0.125)$);
    \draw[locarrow] ($(mainproc.north)+(-0.1,0)$) |- ($(loctrans.east)+(0,-0.1)$);
    \draw[locarrow] ($(loctrans.east)+(0,0.1)$) -| ($(mainproc.north)+(0.1,0)$);
    \draw[darrow] (camera) -| (datafusion.south);
    \draw[darrow] (datafusion) -- (proc);
    \draw[darrow] ($(proc.north)+(0.2,0)$) -- ($(mainproc.south)+(0.2,0)$);
\end{tikzpicture}
        \begin{tikzpicture}
\definecolor{commcolor}{RGB}{33,240,182};
\definecolor{loccolor}{RGB}{194,87,43};
\definecolor{radcolor}{RGB}{199,238,168};

\tikzstyle{block} = [thick,align=center,minimum width=1.5cm,minimum height=1cm,draw,color=black,draw opacity=0.25]
\tikzstyle{antenna} = [black,draw,minimum width=0.5cm,minimum height=0.5cm]
\tikzstyle{darrow} = [line width=1.5pt,-stealth]
\tikzstyle{commarrow} = [darrow,color=commcolor]
\tikzstyle{locarrow} = [darrow,color=loccolor]
\tikzstyle{radarrow} = [darrow,color=radcolor]
    \path (1.1,-1.8) -- ++(4.6,0) -- ++(0,-1.2) -- ++(-4.6,0) -- cycle;
    \draw[black,line width=0.5pt] (1.8,-2) -- ++(4,0) -- ++(0,-1) -- ++(-4,0) -- cycle;
    \draw[commarrow] (2.1,-2.25) -- ++(0.5,0);
    \draw[radarrow] (2.65,-2.25) -- ++(0.5,0);
    \draw[locarrow] (3.2,-2.25) -- ++(0.5,0);
    \draw[commarrow,dashed] (2.1,-2.75) -- ++(0.5,0);
    \draw[radarrow,dashed] (2.65,-2.75) -- ++(0.5,0);
    \draw[locarrow,dashed] (3.2,-2.75) -- ++(0.5,0);
    \node[anchor=west] (legData) at (3.75,-2.25) {Data};
    \node[anchor=west] (legProp) at (3.75,-2.75) {Metadata};
\end{tikzpicture}
        \usetikzlibrary{calc}                   
\usetikzlibrary{shapes}      
\usetikzlibrary{arrows,arrows.meta,decorations.pathmorphing,backgrounds,positioning,fit,petri,patterns}
\usetikzlibrary{shapes.geometric,backgrounds}

\tikzset{
  mybackground/.style={execute at end picture={
        \begin{scope}[on background layer]
          \draw[black!15,fill=gray!5,rounded corners=1ex] ($(current bounding box.south west)+(-0.1,-0.1)$) rectangle ($(current bounding box.north east)+(0.1,0.1)$);
          \node[draw,fill=white,rectangle,rounded corners=0.1,anchor=west,inner sep=2pt,minimum width=4ex,draw opacity=0.25] at ($(current bounding box.north west)+(0.25,0)$) {#1};
        \end{scope}
    }},
}

\begin{tikzpicture}[mybackground={JCSL}]
\definecolor{commcolor}{RGB}{33,240,182};
\definecolor{loccolor}{RGB}{194,87,43};
\definecolor{radcolor}{RGB}{199,238,168};

\tikzstyle{block} = [thick,align=center,minimum width=1.5cm,minimum height=1cm,draw,color=black,draw opacity=0.25]
\tikzstyle{antenna} = [black,draw,minimum width=0.5cm,minimum height=0.5cm]
\tikzstyle{darrow} = [line width=1.5pt,-stealth]
\tikzstyle{commarrow} = [darrow,color=commcolor]
\tikzstyle{locarrow} = [darrow,color=loccolor]
\tikzstyle{radarrow} = [darrow,color=radcolor]
\tikzset{
diagonal fill/.style 2 args={fill=#2, path picture={
    \fill[#1, sharp corners] (path picture bounding box.south west) -|
                         (path picture bounding box.north east) -- cycle;}},
three stripes fill/.style n args={3}{fill=#2, path picture={
    \fill[#1, sharp corners] (path picture bounding box.70) -| (path picture bounding box.200) -- cycle; 
    \fill[#3, sharp corners] (path picture bounding box.250) -| (path picture bounding box.20) -- cycle;}}
}
    
    \coordinate (mma1) at (-2,0.1);
    \coordinate (mma2) at (-2.6,-0.3);
    \coordinate (mma3) at (-2.3,-0.75);

    \node[block,three stripes fill={commcolor!25}{radcolor!25}{loccolor!25},anchor=west] (signalfusion) at (-0.4,0) {JCSL\\Unit};
    \node[block,fill=commcolor!25] (commtrans) at (2.3,1.25) {Communication\\Transceiver};
    \node[block,diagonal fill={radcolor!25}{loccolor!25}] (datafusion) at (2.5,-0.8) {Data\\Fusion};
    
    \node[block,dashed,anchor=west] (camera) at (-0.4,-1.6) {Other\\Sensors};
    \node[block,dashed] (proc) at (5,-0.8) {OA, ATOL\\etc.};

    \node[block,fill=white] (mainproc) at (5,1.25) {Flight\\Controller};

    \draw[black,line width=2pt] ($(mma1)+(0.25,0)$) |- (signalfusion);
    \draw[black,line width=2pt] ($(mma2)+(0.25,0)$) |- (signalfusion);

    \draw[antenna,three stripes fill={commcolor!25}{radcolor!25}{loccolor!25}] (mma1) -- ++(0.5,0) -- ++(0.25,0.25) -- ++(0,0.5) -- ++(-0.5,0) -- ++(-0.25,-0.25) -- ++(0,-0.5) -- ++(0.5,0) -- ++(0,0.5) -- ++(-0.5,0) -- ++(0.5,0) -- ++(0.25,0.25);
    \draw[antenna,three stripes fill={commcolor!25}{radcolor!25}{loccolor!25}] (mma2) -- ++(0.5,0) -- ++(0.25,0.25) -- ++(0,0.5) -- ++(-0.5,0) -- ++(-0.25,-0.25) -- ++(0,-0.5) -- ++(0.5,0) -- ++(0,0.5) -- ++(-0.5,0) -- ++(0.5,0) -- ++(0.25,0.25);
    \draw[antenna,three stripes fill={commcolor!25}{radcolor!25}{loccolor!25}] (mma3) -- ++(0.5,0) -- ++(0.25,0.25) -- ++(0,0.5) -- ++(-0.5,0) -- ++(-0.25,-0.25) -- ++(0,-0.5) -- ++(0.5,0) -- ++(0,0.5) -- ++(-0.5,0) -- ++(0.5,0) -- ++(0.25,0.25);

    \draw[commarrow] ($(commtrans.west)+(0,0.1)$) -| ($(signalfusion.north)+(-0.1,0)$);
    \draw[commarrow,dashed] ($(commtrans.west)+(0,-0.1)$) -| ($(signalfusion.north)+(0.1,0)$);
    \draw[commarrow] ($(signalfusion.east)+(0,0.3)$) -| (commtrans.south);
    \draw[commarrow] ($(mainproc.west)+(0,0.2)$) -- ($(commtrans.east)+(0,0.2)$);
    \draw[commarrow,dashed] (mainproc) -- (commtrans);
    \draw[commarrow] ($(commtrans.east)+(0,-0.2)$) -- ($(mainproc.west)+(0,-0.2)$);
    \draw[radarrow,dashed] ($(mainproc.south)+(-0.2,0)$) |- ($(signalfusion.east)+(0,0.1)$);
    \draw[locarrow,dashed] ($(mainproc.south)+(0,0)$) |- ($(signalfusion.east)+(0,-0.1)$);
    \draw[radarrow] ($(signalfusion.south)+(0.1,0)$) |- ($(datafusion.west)+(0,0.1)$);
    \draw[locarrow] ($(signalfusion.south)+(-0.1,0)$) |- ($(datafusion.west)+(0,-0.1)$);
    \draw[darrow] ($(camera.east)+(0,-0.1)$) -| (datafusion);
    \draw[darrow] (datafusion) -- (proc);
    \draw[darrow] ($(proc.north)+(0.2,0)$) -- ($(mainproc.south)+(0.2,0)$);
\end{tikzpicture}
        \caption{Block diagram of the proposed JCSL system as a part of the communication and localization system for UAVs. Inter-system interference and coupling between the different antenna systems occurring traditionally (top) is avoided by employing the JCSL unit (bottom). Improved data fusion can enhance obstacle avoidance (OA) and automatic take-off and landing (ATOL) processing.
        }
        \label{fig:jcsl block diagram}
    \end{figure}
    Regarding the design of JCSL systems in time and frequency domain, three different approaches with increasing complexity and performance can be differentiated.
    In the simplest form, the radio signals of communication and sensing systems can on one hand be separated in the time domain via a time-multiplexed waveform design, thus reducing the time budget per task and potentially leading to a compromised performance of each one.
    On the other hand, a separation in the frequency domain via a frequency-multiplexed waveform design can be accomplished by utilizing separated frequency bands for each task. 
    This can be costly as the frequency spectrum is a valuable and scarce resource and international regulations restrict the use of it. 
    A more complex approach relies on using one waveform for both approaches in a modified way.
    When a radar signal is simultaneously used for (secondary) communication applications, this can be accomplished by embedding information into the existing radar signal, for instance via additional phase coding. 
    Such an approach is often referred to as \emph{RadCom} as the radar system defines the base for the joint system.
    Preferably, the embedded information should not deteriorate the radar performance in this setup.
    Vice-versa, communication waveforms that are characterized by favorable radar properties, e.g., regarding their autocorrelation properties in time and/or frequency, can also be used for target object detection and localization by means of range/Doppler processing. 
    In accordance with the prior definition, such approaches are referred to as \emph{ComRad} systems.
    In this setup, the communication task is the primary one.
    Fig.~\ref{fig:jcsl block diagram} illustrates the difference between traditional systems and a JCSL system.
    In traditional system designs, each part of the system requires its own system components, including processing, antennas, and RF components.
    Contrarily, a JCSL system designed to fit all requirements and shares a major part of the processing, RF components, and antennas.
    If the same amount of antennas is used, a diversity gain is achievable.
    Vice versa, the required hardware can be reduced, which leads to a reduction of occupied space and weight.
    Hence, the main challenge of the JCSL design is the JCSL unit. 
    In this unit, the data and protocol information of the communication transceiver need to be matched with the requirements of the radar system, which are defined by the flight controller and its needs in terms of sensing performance.
    
    Finally, when JCSL systems are designed bottom up, the waveform can be tailored to meet the requirements of specific tasks. 
    In this context, orthogonal frequency-division multiplexing (OFDM) has been shown to be a good candidate to serve as a modulation scheme for both communication and radar tasks, see for instance \cite{sturm2011waveform} and references therein.
    In conjunction with an appropriate digital single-carrier modulation scheme -- such as M-ary phase-shift keying (PSK) or quadrature amplitude modulation (QAM) -- the orthogonal subcarriers are well suited for signal separation at the receiver and are even robust against Doppler shifts when the subcarrier spacing is chosen appropriately.
    Although being well suited for JCSL applications, OFDM comes with the disadvantage of relatively high peak-to-average power ratios (PAPRs), however. If no special countermeasures are implemented, this requires the transmitter to employ a corresponding power backoff, in order to avoid clipping and non-linear behavior within the power amplifier.
    Yet, reducing the transmit power entails an undesired decrease regarding the maximum achievable radar and communication range.

   Given a pure radar application without simultaneous wireless communication functionality, classic techniques (e.g., Newman phases), may be employed for PAPR reduction, which are typically tailored to real-valued transmit signals \cite{MPVZ19}. 
   To this end, a fixed complex-conjugate symmetry between the OFDM sub-carriers can be established in the complex baseband domain, in order to obtain a real-valued multi-carrier signal, as no random data symbols need to be included at this point. 
   For RadCom applications, however, more advanced PAPR reduction techniques are required, which are able to handle random data symbols mapped onto the individual OFDM sub-carriers (see, e.g., Reference [9] in \cite{mietzner2019dftspread} for an overview). 
   An alternative approach for moderate data-rate communication links could lie in alternative modulation schemes, such as Discrete Fourier-Transform (DFT)-spread OFDM~\cite{mietzner2019dftspread}. 
   Similar to OFDM, a time domain matched filter can be employed for DFT-spread OFDM, in order to maximize the signal-to-noise ratio (SNR) in the range domain. 
   Alternatively, frequency-domain signal processing may be used.

\begin{figure}
		\centering
		\setlength{\figH}{0.5\linewidth}
		\setlength{\figW}{0.8\linewidth}
		\input{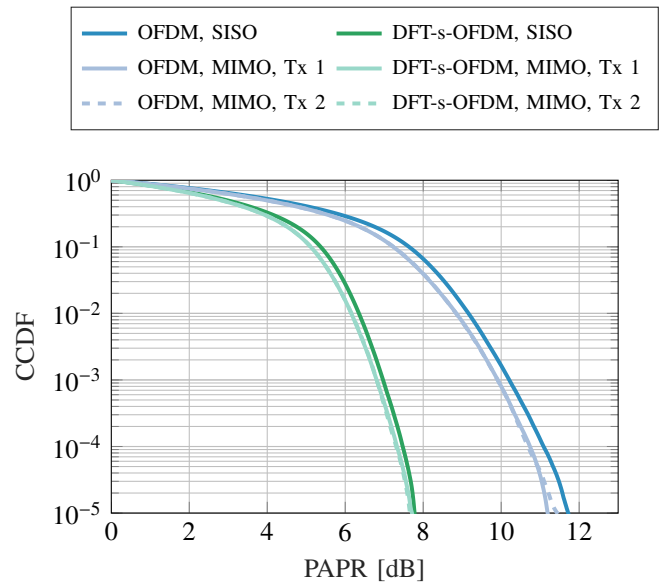}
		\caption{CCDFs for the resulting PAPR values simulated for both a SISO and a MIMO scenario using OFDM and DFT-spread OFDM.}
		\label{fig:papr compared dft ofdm}
	\end{figure}

   A corresponding analysis for DFT-spread OFDM showed a significant reduction in PAPR of several decibel compared to conventional OFDM. 
   An example is illustrated in Fig.~\ref{fig:papr compared dft ofdm} in form of the complementary cumulative distribution functions (CCDFs) of resulting PAPR values. 
   The simulations were performed for a single-input single-output (SISO) and a multiple-input multiple-output (MIMO) configuration with two transmit antennas and interleaved subcarriers. 
   Specifically, 256 sub-carriers with a spacing of $\Delta f=\SI{15\,}{\kilo\hertz}$ have been used in the simulations (the sub-carrier spacing might be further enlarged to improve Doppler resilience). 
   The cyclic prefix length has been chosen to allow for maximum target ranges of $\SI{2\,}{\kilo\meter}$, and an 8-ary PSK modulation was employed, in order to include random data symbols. 
   As can be seen, with DFT-spread OFDM the resulting PAPR-value is below $8\,\mathrm{dB}$ with a probability exceeding $1-10^{-5}$ (corresponding to $99.999\,\%$), while conventional OFDM is associated with significantly larger PAPR values.
   Overall, PAPR-reduced OFDM or DFT-s-OFDM waveforms thus seem to be promising candidates for JCSL applications in the context of future UAM scenarios. 
    
	\section{Miniaturization of Antennas and Application of Multi-Port Antennas for JCSL in Aviation}
	\label{sec:AntennasLUH}
	The demand in efficient use of spectra increases and the miniaturization and automatization of aerial vehicles requests for light-weight but emerging technological solutions.
	
	As shown in Fig.~\ref{fig: PiperAntennas}, currently a large number of antennas is employed on airplanes to fulfill the different needs of the communication and radar systems aboard. 
	By introducing antennas allowing broadband transmission and the use of different antenna patterns at the same time, weight and number of antennas could be reduced.
	This especially holds for the co-design of antennas and JCSL systems and can allow the integration on even smaller and light-weight aerial vehicles.
	The aerial use-case changes some of the requirements during the design process of the antennas.
	First, the antenna system has space and weight constraints.
	Additional weight which has to be lifted increases fuel and power consumption.
	For the same reason, space aboard aircraft usually is limited.
	The chassis of aerial vehicles typically is designed such that it fulfills the requirements for its main task, which is lifting a certain payload.
	Therefore, introducing antennas and radomes can lead to positioning conflicts, which should be taken care of.
	As an example, if communication systems shall be employed next to sensing systems, as typically done, the antennas of one system shadow the antennas and sensors of the other or cause inter-system interference.
	This encourages the design of JCSL especially in aviation, including the design of the antennas.
	Here, M$^3$PAs are of special interest, as they are capable of being designed for wideband and ultra-wideband applications, as shown in \cite{MaMa16,Johannsen2020} and particularly for direction-of-arrival estimation in \cite{GrMa22}.
	The latter antenna has been designed for transponder-based air-to-air communication based on the ACAS~II standard.
	Using the theory of characteristic modes, it has been shown that the symmetry group of a given antenna structure determines the maximum number of uncorrelated ports on the structure. Therefore, a sufficiently symmetric structure is chosen.
	\begin{figure}
		\centering
		\includegraphics[width=\columnwidth]{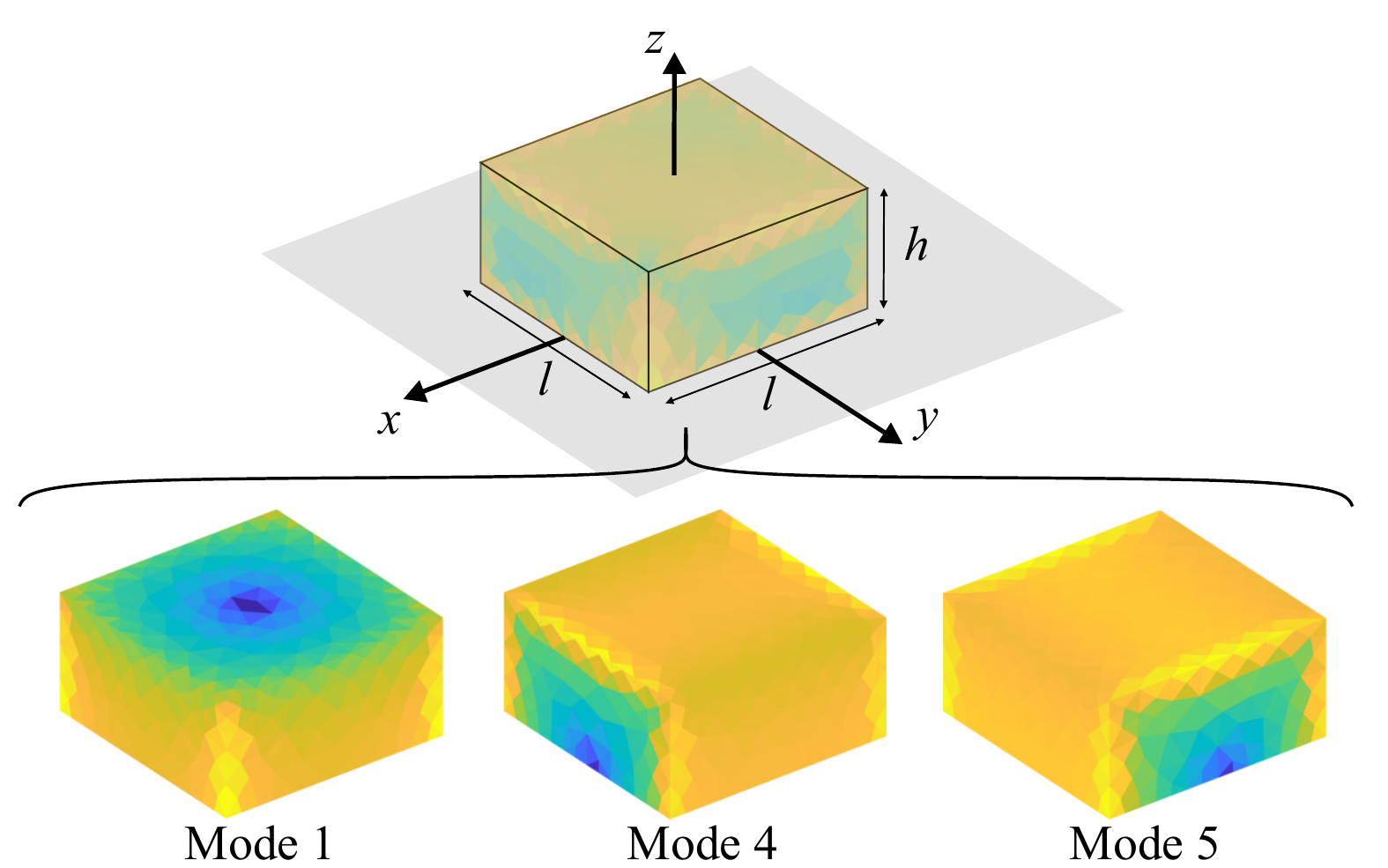}
		\caption{Example shape of a cuboid-shaped antenna prototype as discussed in \cite{GrMa22} with modal current densities.}
		\label{fig:CubicAntenna}
	\end{figure}
    The antenna depicted in Fig.~\ref{fig:CubicAntenna} allows the usage of three independent ports, including the omni-directional monopole mode~1. Note that the limitation to three ports is a design decision rather than being imposed by the shape of the antenna. It is selected to allow a simpler connection to the signal processing.
    Given the discussed scenario of JCSL, the omni-directional monopole mode could be used for communication, like broadcasting information.
    The two additional ports are providing radiation patterns based on orthogonal sets of modes pointing back and forwards, as well as left and right, respectively, which is shown by curves $G_4$ and $G_5$ in Fig.~\ref{fig:Beamforming}.    
	By employing suitable weighting coefficients using all three ports, an improved and steerable directivity can be achieved, as shown by curves $G_{\textrm{max}}$ and $G_{30^{\circ}}$ in Fig.~\ref{fig:Beamforming}.
	In an exemplary ACAS II-based, simple JCSL scenario, this could be used to mask a certain angular region when interrogating other aerial vehicles transponders.
	During reception, both direction-of-arrival and distance can be estimated, based on the received signal and its round trip time.
	The estimation can be used to improve the situational awareness of the system and to possibly avoid collisions.
	
	\begin{figure}[h]
		\centering
		    \begin{tikzpicture}
        \definecolor{clr1a}{RGB}{166,189,219}
        \definecolor{clr1b}{RGB}{43,140,190}
        \definecolor{clr2a}{RGB}{153,216,201}
        \definecolor{clr2b}{RGB}{44,162,95}
        \definecolor{clr3a}{RGB}{251,154,153}
        \definecolor{clr3b}{RGB}{227,26,28}
    
        \begin{axis}[
        ylabel = Gain in dBi,
        xlabel = Angle in degree,
        xmin = 0,
        xmax = 355,
        grid = both,
        ymin = -10,
        ymax = 10,
        scale = 0.8,
        legend columns = 1,
        legend style={at={(1.2,0.5)},nodes=left,draw,fill=white},
        ]
            \addplot[no markers, dashed, black, line width=1.5pt] table [col sep =comma, x=Phi,  y=GWeights] {./tikz/GainOptimization_Cuboid_GNDxyBplot.csv};
            \addplot[no markers, black, line width=1.5pt] table [col sep=comma, x=Phi, y=GWeight30]{./tikz/GainOptimization_Cuboid_GNDxyBplot.csv};
            \addplot[no markers, thick, clr1b, line width=1.5pt] table [col sep =comma, x=Phi,  y=GPortSel1] {./tikz/GainOptimization_Cuboid_GNDxyBplot.csv};
            \addplot[no markers, thick, clr2b,line width=1.5pt] table [col sep =comma, x=Phi,  y=GPortSel2] {./tikz/GainOptimization_Cuboid_GNDxyBplot.csv};
            \addplot[no markers, thick, clr3b,line width=1.5pt] table [col sep =comma, x=Phi,  y=GPortSel3] {./tikz/GainOptimization_Cuboid_GNDxyBplot.csv};
            \addplot+[ycomb,black,line width=1.5pt] plot table[x index=0, y index=1]{datafileB.dat};
            
            \legend{\footnotesize{$G_{\textrm{max}}$},
                    \footnotesize{$G_{\textrm{30}^\circ}$},
                    \footnotesize{$G_{\textrm{1}}$},
                    \footnotesize{$G_{\textrm{4}}$},
                    \footnotesize{$G_{\textrm{5}}$}
                    }
            
        \end{axis}
    \end{tikzpicture}
		\caption{Beamforming performance of a set of three orthogonal modes (1, 4 and 5) of the structure depicted in Fig.~\ref{fig:CubicAntenna}. Gain can be optimized by combining the available radiation characteristics for the full angular range (dashed line). However, when optimizing for certain angles, the maximum of the mainlobe may point to a different direction (black curve, optimized for 30$^\circ$).}
		\label{fig:Beamforming}
	\end{figure}
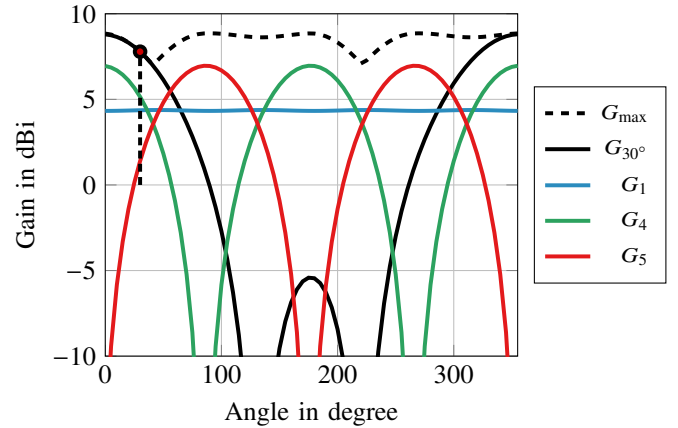
	
	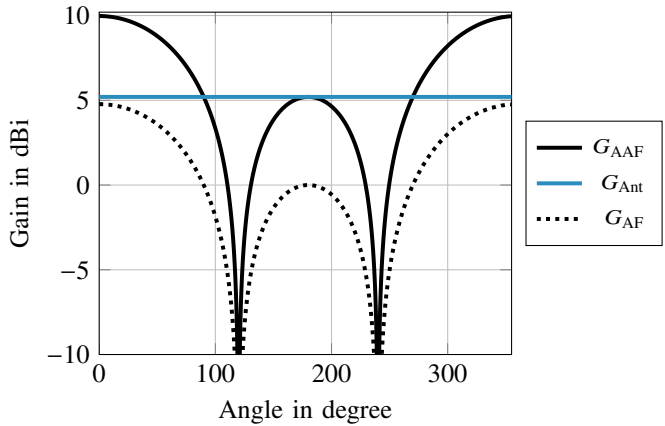
\begin{figure}[h]
	    \centering
	     \begin{tikzpicture}
\definecolor{clr1a}{RGB}{166,189,219}
\definecolor{clr1b}{RGB}{43,140,190}
\definecolor{clr2a}{RGB}{153,216,201}
\definecolor{clr2b}{RGB}{44,162,95}
\definecolor{clr3a}{RGB}{251,154,153}
\definecolor{clr3b}{RGB}{227,26,28}
        \begin{axis}[
        ylabel = Gain in dBi,
        xlabel = Angle in degree,
        xmin = 0,
        xmax = 355,
        grid = both,
        ymin = -10,
        ymax = 10.2,
        scale = 0.8,
        legend columns = 1,
        legend style={at={(1.2,0.5)},nodes=left,draw,fill=white},
        ]
            \addplot[domain=0.1:355,samples=1000,black,line width=1.5pt] {(10*log10(abs(sin(3*x/2)/sin(x/2)))+5.2)};
            \addplot[domain=0.1:355, color=clr1b,line width=1.5pt]{5.2};
        \addplot[domain=0.1:355,samples=1000,color=black,dotted,line width=1.5pt] {(10*log10(abs(sin(3*x/2)/sin(x/2))))};
            
        \legend{\footnotesize{$G_{\textrm{AAF}}$},
                    \footnotesize{$G_{\textrm{Ant}}$},
                    \footnotesize{$G_{\textrm{AF}}$}}
                    
            
        \end{axis}
    \end{tikzpicture}
	    \caption{
	    Beamforming performance of a uniformly fed three element array, using a spacing of half a wavelength.
	    The curves show the antenna and array factor ($G_{\textrm{AAF}}$, array factor plus gain of element), gain of an example monopole antenna ($G_{\textrm{Ant}}$, assumed to be 5.2~dBi), and array factor ($G_{\textrm{AF}}$). These analytical results show that the achievable gain using an antenna array and employing monopole antennas can increase the gain compared to the proposed M$^3$PA but at the cost of an increased sidelobe level.}
	    \label{fig:ThreeElementArray}
	\end{figure}

    In Fig.~\ref{fig:ThreeElementArray}, the beamforming performance in terms of the antenna-and-array factor (AAF) of a three element monopole array is shown. 
    The AAF combines the antenna gain of the employed antenna elements in the array with the processing gain achieved by the structure of the antenna array.
    The assumed ideal quarter-wavelength monopole antenna gain is 5.2~dBi and the distance of the antennas is half a wavelength. 
    The space occupied by the antenna array is similar to the space occupied by the M$^3$PA shown in Fig.~\ref{fig:CubicAntenna}.
    While a slightly larger AAF can be achieved by the array, the sidelobe level is -5~dB, compared to -14~dB using the M$^3$PA. 
    
    These results could find their initial application in an improved version or new generation of interrogators.
    In this scenario, an interrogator equipped with an M$^3$PA is capable of performing a directive interrogation of other vehicles, due to the improved side lobe level.
    Since M$^3$PAs can also be used for localization estimation, transponder-based collision avoidance could be improved by means of additional information.

	\section{Conclusion}
	\label{sec:Conclusion}
	In this paper, JCSL has been discussed as an emerging technology towards challenges and goals of the upcoming trends in urban air mobility.
	It was shown that multi-mode multi-port antennas allow improved performance compared to traditional antenna-array configurations and increase the available number of ports. 
	Hence, this advanced antenna type is a promising candidate for the design of JCSL systems. 
	Furthermore, joint waveform design for JCSL is regarded as an enabling technology for developing highly integrated, multi-functional, and light-weight on-board RF systems meeting the requirements of confined installation spaces within future UAVs. 
	Therefore, the key technologies addressed in this article are expected to offer a significant contribution when elevating autonomous driving to the third dimension.

	\section*{Acknowledgment}
	This project is part of the Master360 program under research grant 20D1905, funded by the German Federal Ministry for Economic Affairs and Climate Action (BMWK).
        Thanks to Nils Segebrecht for providing a picture of his Piper.
        
	The cooperative work with f.u.n.k.e. GmbH, Ulm, Germany, and Dr. Askold Meusling (project leader ``Master360''), Airbus Defence and Space GmbH, Taufkirchen, Germany, is highly appreciated.

	\bibliographystyle{IEEEtran}
	\bibliography{IEEEabrv,ICTabrv.bib,literature.bib}

\begin{thebibliography}{10}
\providecommand{\url}[1]{#1}
\csname url@samestyle\endcsname
\providecommand{\newblock}{\relax}
\providecommand{\bibinfo}[2]{#2}
\providecommand{\BIBentrySTDinterwordspacing}{\spaceskip=0pt\relax}
\providecommand{\BIBentryALTinterwordstretchfactor}{4}
\providecommand{\BIBentryALTinterwordspacing}{\spaceskip=\fontdimen2\font plus
\BIBentryALTinterwordstretchfactor\fontdimen3\font minus
  \fontdimen4\font\relax}
\providecommand{\BIBforeignlanguage}[2]{{%
\expandafter\ifx\csname l@#1\endcsname\relax
\typeout{** WARNING: IEEEtran.bst: No hyphenation pattern has been}%
\typeout{** loaded for the language `#1'. Using the pattern for}%
\typeout{** the default language instead.}%
\else
\language=\csname l@#1\endcsname
\fi
#2}}
\providecommand{\BIBdecl}{\relax}
\BIBdecl

\bibitem{NLALO13}
N.~Neji, R.~D. Lacerda, A.~Azoulay, T.~Letertre, and O.~Outtier, ``Survey on
  the future aeronautical communication system and its development for
  continental communications,'' \emph{{IEEE} Trans. Veh. Technol.}, vol.~62,
  no.~1, pp. 182 -- 191, 2013.

\bibitem{PTSTJ18}
D.~S. Ponchak, F.~L. Templin, G.~Sheffield, P.~Taboso, and R.~Jain, ``An
  implementation analysis of communications, navigation, and surveillance
  ({CNS}) technologies for unmanned air systems ({UAS}),'' in \emph{2018
  IEEE/AIAA 37th Digital Avionics Systems Conference ({DASC})}, 2018, pp.
  1--10.

\bibitem{BDOACS21}
A.~Baltaci, E.~Dinc, M.~Ozger, A.~Alabbasi, C.~Cavdar, and D.~Schupke, ``A
  survey of wireless networks for future aerial communications ({FACOM}),''
  \emph{{IEEE} Commun. Surveys Tuts.}, vol.~23, no.~4, pp. 2833--2884, 2021.

\bibitem{ED275}
EUROCAE, ``Minimum operational performance standards for airborne collision
  avoidance system {X}u ({ACAS X}u),'' \emph{ED 275}, vol.~1, Dec. 2020.

\bibitem{ED73C}
{The European Organisation for Civil Aviation Equipment}, ``Minimum operational
  performance specification for secondary surveillance radar mode {S}
  transponders,'' \emph{ED-73C}, May 2008.

\bibitem{Mar20}
A.~Marsh, ``In {World War I}, {British} biplanes had wireless phones in the
  cockpit,'' \emph{{IEEE} Spectr.}, Mar. 2020,
  https://spectrum.ieee.org/in-world-war-i-british-biplanes-had-wireless-phones-in-cockpit,
  Retrieved 2022/03/21.

\bibitem{MaJe16}
G.~Manfredi and Y.~Jestin, ``An introduction to {ACAS X}u and the challenges
  ahead,'' in \emph{2016 IEEE/AIAA 35th Digital Avionics Systems Conference
  (DASC)}, Sep. 2016, pp. 1--9.

\bibitem{CaPeAl18}
X.~Cao, P.~Yang, M.~Alzenad, X.~Xi, D.~Wu, and H.~Yanikomeroglu, ``Airborne
  communication networks: A survey,'' \emph{{IEEE} J. Sel. Areas Commun.},
  vol.~36, no.~9, pp. 1907--1926, 2018.

\bibitem{Har21}
M.~Harris, ``{FAA} files reveal a surprising threat to airline safety: the
  {U.S.} military's {GPS} tests,'' \emph{{IEEE} Spectr.}, Jan. 2021,
  https://spectrum.ieee.org/faa-files-reveal-a-surprising-threat-to-airline-safety-the-us-militarys-gps-tests,
  Retrieved 2022/03/21.

\bibitem{sturm2011waveform}
C.~Sturm and W.~Wiesbeck, ``Waveform design and signal processing aspects for
  fusion of wireless communications and radar sensing,'' \emph{Proceedings of
  the IEEE}, vol.~99, no.~7, pp. 1236--1259, 2011.

\bibitem{MPVZ19}
T.~Multerer, U.~Prechtel, M.~Vossiek, and V.~Ziegler, ``Systematic phase
  correction for direction-of-arrival estimation in spectrally interleaved
  {OFDM MIMO} radar,'' \emph{{IEEE} Trans. Microw. Theory Techn.}, vol.~67,
  no.~11, pp. 4570 -- 4577, 2019.

\bibitem{mietzner2019dftspread}
J.~Mietzner, ``{DFT}-spread {OFDM} {MIMO}-radar – an alternative for reduced
  crest factors,'' in \emph{2019 20th International Radar Symposium (IRS)},
  2019, pp. 1--10.

\bibitem{MaMa16}
D.~Manteuffel and R.~Martens, ``Compact multimode multielement antenna for
  indoor {UWB} massive {MIMO},'' \emph{{IEEE} Trans. Antennas Propag.},
  vol.~64, no.~7, pp. 2689--2697, Jul. 2016.

\bibitem{Johannsen2020}
N.~L. Johannsen, N.~Peitzmeier, P.~A. Hoeher, and D.~Manteuffel, ``On the
  feasibility of multi-mode antennas in {UWB} and {IoT} applications below {10
  GHz},'' \emph{IEEE Communications Magazine}, vol.~58, no.~3, pp. 69--75,
  2020.

\bibitem{GrMa22}
L.~Grundmann and D.~Manteuffel, ``Selecting characteristic modes in multi-mode
  direction finding antenna design by using reconstructed incident fields,'' in
  \emph{2022 16th European Conference on Antennas and Propagation (EuCAP)},
  2022, pp. 1--5.

\end{thebibliography}
	
	\begin{IEEEbiographynophoto}{Nils Lennart Johannsen}
	    (nj@tf.uni-kiel.de) received his B.Eng. from Hamburg University of Applied Sciences, Germany, in 2016 and an M.Sc. degree in electrical engineering and information technology from Kiel University, Germany, in 2018. He is currently pursuing a Ph.D. degree with the Chair of Information and Coding Theory, Kiel University, as a research and teaching assistant. His research interests include massive MIMO systems and baseband signal processing for multi-mode antennas.
	\end{IEEEbiographynophoto}
	
	\begin{IEEEbiographynophoto}{Max Schurwanz}
	    (max.schurwanz@haw-hamburg.de) received the M.Sc. degree in electrical engineering from Kiel University, Germany, in 2020. He is currently pursuing the Dr.-Ing. (Ph.D.) degree at University of Applied Sciences, Hamburg and Kiel University. His current research interests include radar signal processing and investigation of RF-systems co-existence.
	\end{IEEEbiographynophoto}
	
	\begin{IEEEbiographynophoto}{Lukas Grundmann}
	    (grundmann@imw.uni-hannover.de) received the B.Sc. and M.Sc. degrees in electrical engineering from Leibniz University Hannover, Hannover, Germany, in 2017 and 2019, respectively. He is currently a Research Assistant with the Institute of Microwave and Wireless Systems, Leibniz University Hannover. His current research focuses on modal expansion techniques, such as spherical wave functions and characteristic modes, and their applications to antenna development, in particular aerial direction finding. 
	\end{IEEEbiographynophoto}
	
	\begin{IEEEbiographynophoto}{Jan Mietzner} [SM] (jan.mietzner@haw-hamburg.de)
	    received the Ph.D. degree (Hons.) in electrical and information engineering from Kiel University, Germany, in 2006. From 2007 to 2008, he was a Postdoctoral Research Fellow at The University of British Columbia, Vancouver, BC, Canada. In 2009, he joined Airbus DS (now Hensoldt), Ulm, Germany, and worked in the area of jamming and radar systems (especially on MIMO radar). In September 2017, he became a Professor of communications engineering at the Hamburg University of Applied Sciences (HAW). 
	\end{IEEEbiographynophoto}
	
	\begin{IEEEbiographynophoto}{Dirk Manteuffel}
	    (manteuffel@hft.uni-hannover.de) received his Dipl.Ing. and Dr.Ing. degrees in electrical engineering from the University of Duisburg-Essen, Germany, in 1998 and 2002, respectively. From 1998 to 2009, he was with IMST, Kamp-Lint-fort, Germany. From 2009 to 2016, he was a full professor of wireless communications at Christian-Albrechts-University, Kiel, Germany. Since June 2016, he has been a full professor and the director of the Institute of Microwave and Wireless Systems, Leibniz University Hannover, Germany.
	\end{IEEEbiographynophoto}

	\begin{IEEEbiographynophoto}{Peter Adam Hoeher}
	    [F'14] (ph@tf.uni-kiel.de) received his Dipl. Ing. degree from RWTH Aachen University, Germany, in 1986 and his Dr.Ing. degree from the University of Kaiserslautern in 1990, both in electrical engineering. From 1986 to 1998, he was with the German Aerospace Center, Oberpfaffenhofen. From 1991 to 1992, he was on leave at AT\&T Bell Laboratories, Murray Hill, New Jersey. Since 1998 he has been a full professor of electrical and information engineering at Kiel University, Germany.
	\end{IEEEbiographynophoto}
\end{document}